\documentclass{elsart}

\usepackage{graphicx}
\usepackage{amssymb}

\begin{document}
	\begin{frontmatter}

		\title{Heisenberg antiferromagnets with uniaxial exchange and cubic
		anisotropies in a field}
	
		\author{G. Bannasch and}
		\author{W.~Selke}
		
		\address{Institut f\"ur Theoretische Physik, RWTH Aachen, 52056 Aachen, Germany}
	
		\begin{abstract}
			Classical Heisenberg antiferromagnets with uniaxial exchange
			anisotropy and a cubic anisotropy term in a field on simple cubic lattices are
			studied with the help of ground state considerations and extensive
			Monte Carlo simulations. Especially, we analyze the role of non--collinear
			structures of biconical type occurring in addition to
			the well--known antiferromagnetic and spin--flop structures. Pertinent
			phase diagrams are determined, and compared to previous findings.
		\end{abstract}
		
	
	\end{frontmatter}
\section{Introduction}

Uniaxially anisotropic Heisenberg antiferromagnets in a
magnetic field have been
studied quite extensively in the past, both experimentally
and theoretically. Typically, they display, at low
temperatures, the antiferromagnetic phase and, when increasing
the field, the spin--flop phase \cite{Neel}. A prototypical model
describing these phases is the Heisenberg model with a
uniaxial exchange anisotropy, the XXZ model

 \begin{equation}
  {\cal H}_{\mathrm{XXZ}} = J \sum\limits_{i,j}
  \left[ \, \Delta (S_i^x S_j^x + S_i^y S_j^y) + S_i^z S_j^z \, \right]
  \; - \; H \sum\limits_{i} S_i^z
\end{equation}

where $J$ is the exchange coupling between classical
spins, $(S_{i(j)}^x,S_{i(j)}^y,S_{i(j)}^z)$, of length one
at neighboring sites, $i$ and
$j$, of a simple cubic lattice, $\Delta$ is the exchange
anisotropy, $1 > \Delta > 0$, and $H$ is the applied
magnetic field along the easy axis,
the $z$--axis. 

\begin{figure}[b]
  \includegraphics[width=0.9\textwidth]{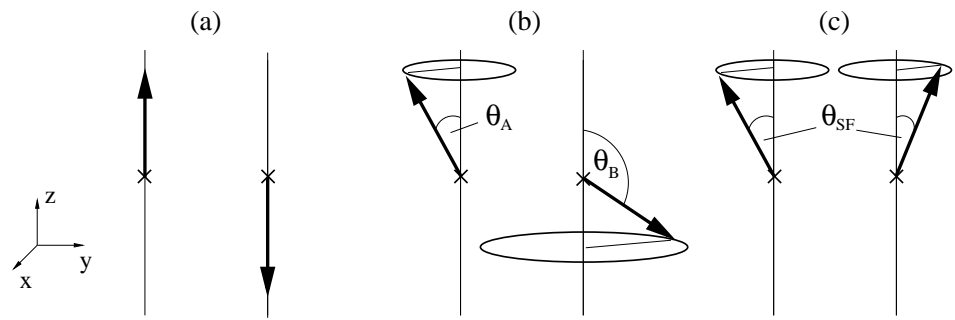}
  \caption{Spin orientations on neighboring sites showing
     antiferromagnetic (a), biconical (b), and spin--flop (c) ground
     state structures of the XXZ model.
    }
  \label{eps1}
\end{figure}

The phase diagram of the model has been studied several years
ago, using mean--field theory \cite{Gorter}, Monte Carlo
simulations \cite{LanBin}, and high temperature series
expansions \cite{Mourit}, suggesting the transition between the
antiferromagnetic (AF) and spin--flop (SF) phases to be of first order
and the
boundaries of the paramagnetic phase to the AF
and SF phases to be continuous transitions in the Ising
and XY universality classes. Based on renormalization group analysis
in one loop order, a bicritical point in the Heisenberg 
universality class has been proposed, at which the
three different phases meet \cite{FN,KNF}. However, this
scenario has been scrutinized when doing
renormalization group calculations in high loop order \cite{CPV}, where the
bicritical point is found to be unstable against a 'tetracritical biconical
point' \cite{KNF}, which, in turn, may be unstable
towards transitions of first order in the vicinity of
the multicritical point of the three
phases. The seemingly conflicting descriptions may be reconciled
by a new renormalization group analysis in two loop order \cite{Folk}.

As has been noted very recently \cite{HWS,HS1,HS2}, not only
AF and SF phases, but also 'biconical' (BC) structures, see
Fig. 1, may play an important role in the XXZ model. Indeed, such
BC structures are degenerate ground states at the critical field 
separating AF and SF configurations at zero temperature. For the
XXZ model on a square lattice, these degenerate BC fluctuations
seem to lead to a narrow disordered phase intervening between
the AF and SF phases at low temperatures \cite{HWS,Zhou1,HSL}. The
importance of BC structures
for the three--dimensional XXZ antiferromagnet, where they are
also present as degenerate ground states at the
special field, had not been
studied in any detail so far.  

Biconical structures may be ground states
even in a finite range of magnetic fields, giving then rise
to an ordered BC phase at low temperatures, when
introducing in the XXZ model further anisotropy
terms or longer--range exchange interactions, as it is known
for many years \cite{MT,LF}. This
feature has been confirmed in recent simulations when adding
a quadratic single--ion 
anisotropy to the XXZ model on the square lattice \cite{HS1,HS2}. The
related lowest--order single--ion term of cubic symmetry may
be written in the form \cite{Kef,Aha}
\begin{equation}
  {\cal H}_{\mathrm{CA}} =  F \sum\limits_{i} \left[(S_i^x)^4 + (S_i^y)^4+
(S_i^z)^4 \right]
\end{equation}
where $F$ denotes the strength of the 'cubic anisotropy'. The sign of
$F$ determines whether the spins tend to align along the cubic
axes, for $F < 0$, or, for $F > 0$, in the diagonal directions
of the lattice. Because
of these tendencies, the BC structures show no
full rotational invariance
in the $xy$--plane perpendicular to the easy axis, the $z$--axis, in contrast
to the XXZ case. 

When an ordered BC phase exists at low temperatures, for example
due to the cubic anisotropy, intricate multicritical behavior may show
up, including a tetracritical biconical point, at which the AF, SF,
BC, and paramagnetic phases meet, as
has been discussed before \cite{LF,Weg,BA,Aha2}, applying mean-field theory
and renormalization group arguments.

Experimentally, many antiferromagnets with uniaxial anisotropy
have been investigated, quasi two--dimensional magnets \cite{DJ,T,BS} as
well as three--dimensional magnets such as
GdAlO$_3$, NiCl$_2$4H$_2$O, MnF$_2$ or Mn$_2$(Si,Ge)S$_4$
\cite{RG,BO,FK,OU,OP}. While we shall deal here with the theoretical
analysis of the
models, results may turn out to be useful for interpreting
specific experiments in future work.   

The aim of our paper is to study, especially, the role of
biconical, non--collinear 
structures in three--dimensional classical Heisenberg antiferromagnets
with uniaxial exchange anisotropy and cubic anisotropy. Both
ground state considerations and
Monte Carlo techniques are applied. 

\begin{figure}[b]
  \includegraphics[width=.7\textwidth]{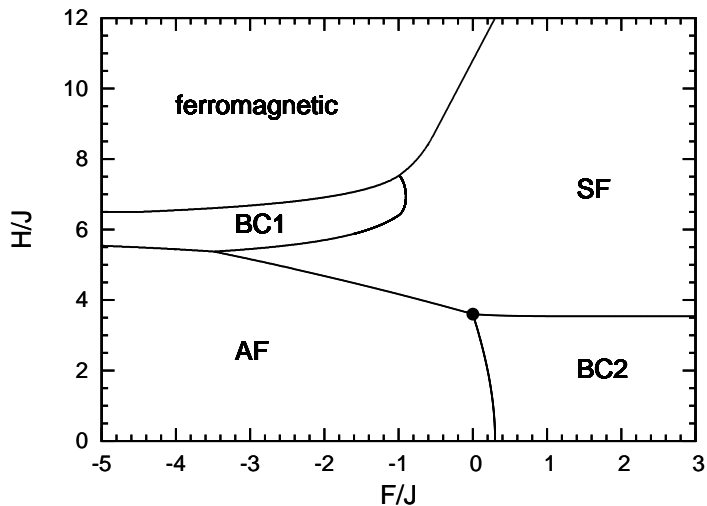}
  \caption{Ground states in the $(F,H)$--plane for the full
    Hamiltonian, ${\cal H}_{\mathrm{f}}$= ${\cal H}_{\mathrm{XXZ}}$ + 
    ${\cal H}_{\mathrm{CA}}$, with exchange anisotropy
    $\Delta$= 0.8. The full circle denotes the highly degenerate point
    in the XXZ model.
    }
  \label{eps2}
\end{figure}

The paper is organized as follows: First, results of 
ground state calculations are presented, and then Monte
Carlo findings on thermal properties and phase diagrams 
will be discussed, for the XXZ model and its extension. The
article will be concluded by a
short summary.

\section{Ground state properties}

The ground states of the full Hamiltonian, 
 ${\cal H}_{\mathrm{f}}$= ${\cal H}_{\mathrm{XXZ}}$ + ${\cal
   H}_{\mathrm{CA}}$, eqs. (1) and (2), may be determined by
 minimization of the energy, e.g., with respect to the
azimuthal angle, $\phi$, i.e. the angle between the projection of
the spin vector in the $xy$--plane and the $x$--axis, and
the $z$--component of the spin vector 
\cite{MT,Ban}. As usual, we consider spin structures with two sublattices, $A$
and $B$, where neighboring sites on the cubic lattice belong to different
sublattices. In general, the minimization may be easily
done numerically.

In Fig. 2, resulting ground states in the $(F,H)$--plane are
depicted, setting the exchange anisotropy $\Delta$ equal
to 0.8, as before \cite{LanBin,HWS,Zhou1}.
 
\begin{figure}[b]
  \includegraphics[width=0.9\textwidth]{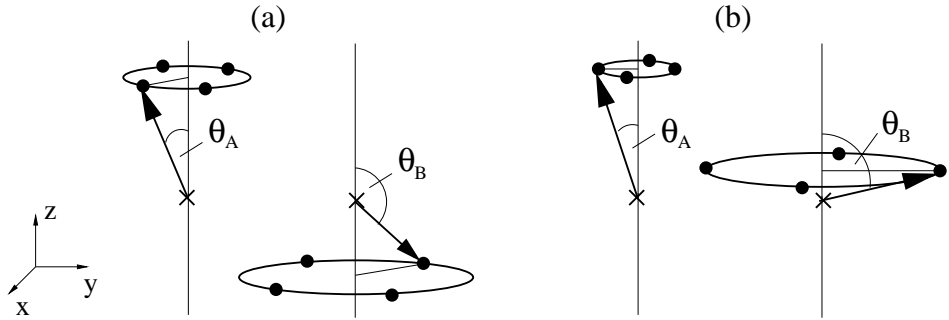}
  \caption{Sketch of discretized biconical structures for
    antiferromagnets with cubic anisotropy, ${\cal H}_{\mathrm{f}}$,
    in the case of (a) $F >0$, BC2, and (b)
    $F<0$, BC1.
    }
  \label{eps3}
\end{figure}

In the case of a vanishing cubic anisotropy, $F= 0$, i.e. for the
XXZ antiferromagnet, one encounters, when
increasing the field, AF, SF, and ferromagnetic ground state
configurations. In complete analogy to the 
model on a square lattice, the ground state is highly degenerate
in BC structures at the critical field $H_c$ separating the
AF and SF structures, see Figs. 1 and 2. For cubic
lattices, one has $H_c= 6J \sqrt{1- \Delta ^2}$. The
non--collinear BC spin structures may be characterized by
two different polar or tilt angles of the spin
vectors on the two sublattices $A$ and $B$, $\Theta_A$ and
$\Theta_B$, as shown in Fig. 1(b). Obviously, for the SF structures,
the angles are identical, $\Theta_A= \Theta_B= \Theta_{SF}$, depicted
in Fig. 1(c). At $H_c$, the two tilt angles of the degenerate BC structures
are interrelated by

\begin{equation}
 \Theta_B = \arccos \left( \frac{ \sqrt{1-\Delta^2} \; - \; \cos\Theta_A }{ 1 \; - \; \sqrt{1-\Delta^2} \cos\Theta_A } \right)
\end{equation}

interpolating continuously between the AF and SF configurations, when
varying one of the tilt angles \cite{HWS,HS1}. Of
course, the rotational invariance of the spin components in
the $xy$--plane leads to an additional degeneracy in the BC and
SF configurations. 
 
Let us now consider positive values of $F$, $F>0$. For
vanishing field, $H$= 0, and $F/J >0.3$, see Fig. 2, the simple 
AF structure is replaced by a tilted antiferromagnetic (TAF) 
configuration. Spins on the
two sublattices point still in opposite directions, but not along the 
easy axis. Indeed, the spins are oriented more and more towards
diagonals of the lattice with increasing positive
cubic anisotropy $F$ \cite
{Ban}. Applying, for $F >0$, an external field, BC structures
may occur, in which neighboring spins have different tilt
angles, $\Theta_A$ and $\Theta_B$. In contrast to the
XXZ case, there is no full rotational symmetry in the
azimuthal angles, $\phi$, of the spins. Instead, for $F>0$, the spins order
along the diagonals of the lattice, and
$\phi$ is discretized, taking the
values, $(2n+1)\pi/4$, where $n$ is an
integer. The $xy$--spin components are ordered
antiferromagnetically, $\phi_B$= $\phi_A +\pi$. The resulting 
discretized biconical structure, BC2, is
sketched in Fig. 3(a). Obviously, the SF configurations
are discretized in the azimuthal angle in the same way
as the BC structures. When fixing $F$ and varying the
field $H$, the two tilt angles, $\Theta_A$ and
$\Theta_B$, change continuosly when going from the AF (or TAF) to the BC2
structures. In the BC2 region, the tilt angles, $\Theta_A$ and
$\Theta_B$, vary continuously
as well. On the other hand, they seem to jump when going from
the BC2 to the SF region.  Note that the tilt angle of the SF
structure, $\Theta_{SF}$, at
the border to the BC2 region, see Fig. 2, depends only weakly
on the strength $F$ of
the cubic anisotropy.

For $F <0$, the azimuthal
angles of the spins, both for biconical and SF structures, take the
values $\phi =n\pi/2$, $n$ being an integer,
reflecting the fact that the cubic anisotropy now favors alignment
of the spins along the cubic axes. The
$xy$--spin components are
ordered antiferromagnetically, $\phi_B$= $\phi_A +\pi$, as
for $F>0$. Biconical ground states occur for sufficiently
strong negative values of $F$, as shown in Fig. 2. The resulting BC1 structure
is shown in Fig. 3(b). The tilt
angles $\Theta_A$ and $\Theta_B$ change continuously in
the BC1 region, when varying the cubic
anisotropy $F$ and the field $H$. At the boundaries of the
BC1 region to the SF, AF, and ferromagnetic
regions the tilt angles seem to jump, except between
the BC1 and SF regions near the triple point of these two and the
ferromagnetic regions, see Fig. 2.

\section{Monte Carlo simulations}

\begin{figure}[b]
  \includegraphics[width=.7\textwidth]{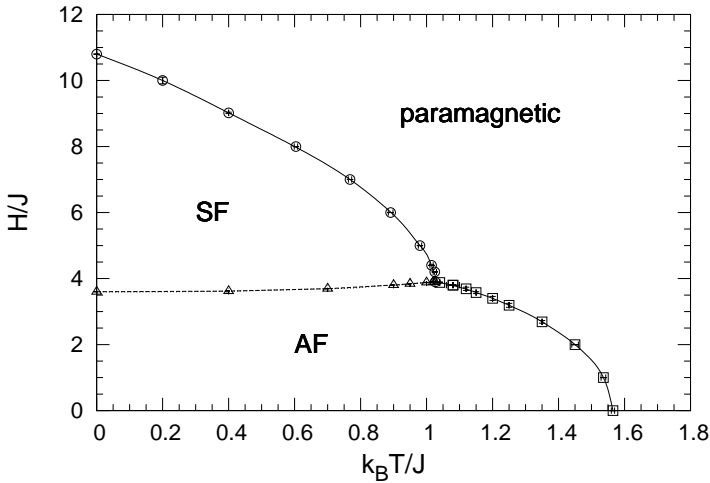}
  \caption{Phase diagram of the XXZ model on a cubic lattice
    with exchange anisotropy $\Delta= 0.8$, as obtained from
    Monte Carlo simulations described in the text.
    }
  \label{eps4}
\end{figure}

In our simulations of thermal properties of the XXZ and the full
model, ${\cal H}_{\mathrm{f}}$, we
applied the Metropolis algorithm with single--spin--flips. Lattices
of $L^3$ sites, employing periodic boundary conditions, with
$L$ ranging from 4 to 32, allowed to do useful finite--size analyses. To obtain
thermal averages, we performed, at fixed model parameters and
temperature, several independent runs, with distinct random numbers, each
run consisting of, at least, $10^7$ Monte Carlo steps per site for the
larger systems. 

We recorded standard thermodynamic quantities such as the specific
heat, $C$, the (absolute) total
magnetization, $m$, longitudinal, $m^z_{st}$, and
transverse, $m^{xy}_{st}$, staggered magnetizations, being the
order parameters in the AF and SF phases, as well as the
corresponding susceptibilities, $\chi$, $\chi^z_{st}$, and 
$\chi^{xy}_{st}$. In addition, we computed the Binder
cumulants \cite{Bin} of the 
two different order parameters, $U_z$ and $U_{xy}$, and
related histograms. To
gain microscopic insights, especially, on BC
structures, we also recorded  probability functions of the
tilt angles, such as the probability $p_2(\Theta_A,\Theta_B)$ 
for finding the two angles, $\Theta_A$ and $\Theta_B$, at
neighboring sites, and the probability $p(\Theta)$ for
encountering the tilt angle $\Theta$ \cite{HWS,HS1,HS2}.

\begin{figure}[b]
  \includegraphics[width=.6\textwidth]{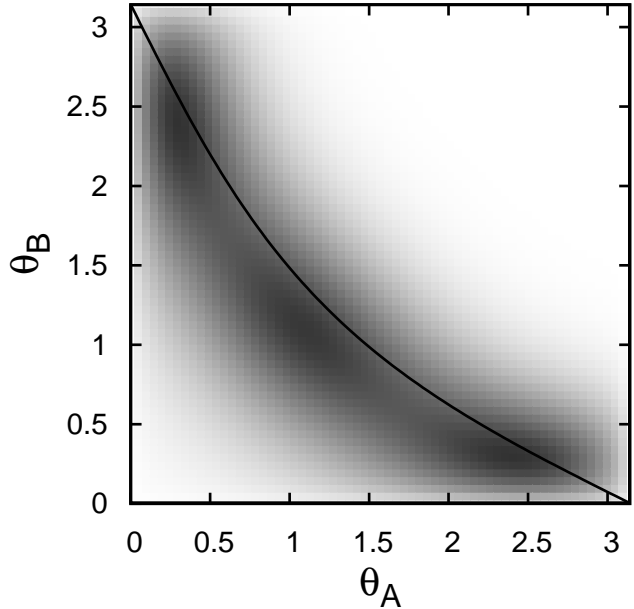}
  \caption{Grayscale representation of the probability $p_2$ for
           finding the tilt angles $\Theta_A$
           and $\Theta_B$ at neighboring sites in the XXZ model with
           $\Delta= 0.8$ in the vicinity of the transition between
           the AF and SF phases, at $k_BT/J= 0.7$ and $H/J=
           3.6925$. The solid line corresponds to eq. (3). The 
           system size is $L= 24$.
           }
  \label{eps5}
\end{figure}

Let us first consider the XXZ model on a cubic lattice, eq. (1). Its
phase diagram has been determined, using Monte Carlo techniques,
already some time ago \cite{LanBin}. Certainly, in present simulations
the accuracy has been improved significantly. The resulting
phase diagram is depicted in Fig.4, where the
phase boundaries have been estimated by standard finite--size
extrapolations \cite{Bar} for thermodynamic quantities and
the Binder cumulant. The phase boundary
lines deviate somewhat from the ones estimated before \cite{LanBin}, which
we attribute to the improved statistics of the present study.

We also identified the type of transitions and, in
the case of continuous transitions, their
universality classes, by monitoring, e.g., the finite
size--dependences of the peak heights of the susceptibilities
and specific heat, to estimate critical exponents, see below.

In agreement with previous suggestions \cite{LanBin,HSL}, the
transition between the AF and SF phases is found to be
of first order, with, for instance, the maximal 
staggered susceptibilities increasing with system size
proportionally to $L^3$, characteristic for such
transitions \cite{VR}. Moreover, as a remainder of the
degeneracy of the ground state at $H_c$, biconical structures prevail
near that transition at low temperatures, as one may
easily observe in $p_2$, see Fig. 5. As included in the
figure, the dominant BC
fluctuations are quite close to those expected
from the degenerate ground states, eq. (3). Note that $p_2$
displays simultaneously local maxima at positions belonging to
the AF and SF structures, as shown in Fig. 5. The
phenomenon gets more pronounced when increasing
the system size, and it reflects coexistence of the AF and SF
phases at a first--order transition. This behavior is in
marked contrast to that of the XXZ antiferromagnet on
a square lattice, where the BC fluctuations seem to lead to
a disordered phase between the AF and SF phases \cite{HWS,HS1,Zhou1,HSL}. 

The transition from the paramagnetic phase to the AF phase
is found to be in the Ising universality class, while the transition
from the paramagnetic phase to the SF phase is found to belong
to the XY universality class. For instance, we estimated
canonical \cite{Bar} asymptotic critical exponents
describing the size dependence
of the peak height of the staggered
susceptibilties, $(\gamma/\nu)^z= d(\ln \chi^z)/d(\ln L)$,
 $L \rightarrow \infty$, and analogously for the transverse
susceptibility, see Fig. 6. Our estimates 
agree nicely with the known, accurate values obtained from
renormalization group calculations in high loop order for
both universality classes \cite{PV}. Similarily, our estimates \cite{Ban}
for the critical Binder cumulants $U^*_z$ and $U^*_{xy}$ at the two types
of transitions are close to those obtained for the three--dimensional
nearest--neighbor Ising and XY models \cite{Ha1,Ha2}. Note, that
the critical Binder cumulant
may change, within a given universality class, due to, e.g., spatially
anisotropic interactions \cite{CD,SeSh}. This possible pitfall
seems to play no role in the XXZ model. 

\begin{figure}[b]
  \includegraphics[width=.7\textwidth]{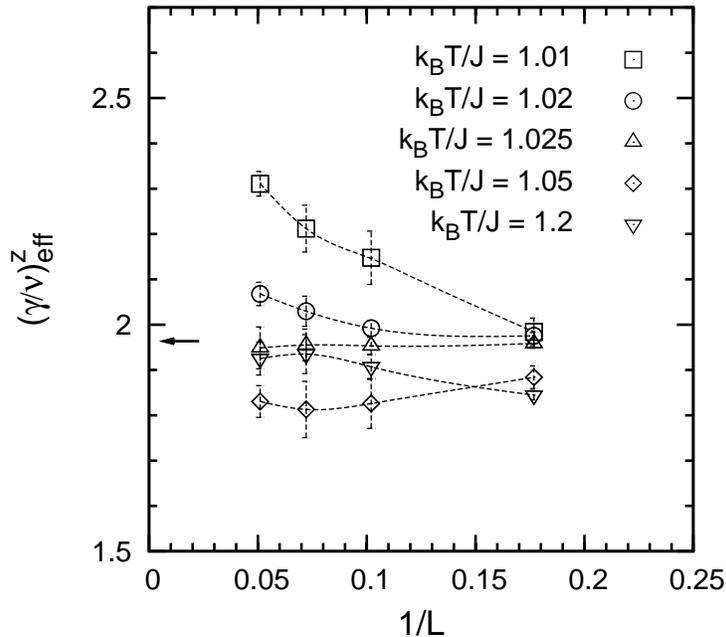}
  \caption{Effective
    critical exponent $(\gamma/\nu)^z_{eff}(L)= d \ln \chi^z/d \ln L$ 
    of the staggered longitudinal susceptibility, to locate the
    multicritical point in the XXZ model. The arrow marks the
    asymptotic critical exponent in the Ising universality class \cite{PV}.
           }
  \label{eps6}
\end{figure}

Analyzing critical exponents
and critical Binder cumulants, we locate the multicritical
point at $k_BT/J= 1.025 \pm 0.015$ and 
$H/J= 3.90 \pm 0.03$, improving the previous
estimate \cite{LanBin}. To obtain the estimate, we
proceed as exemplified for
$\chi^z_{st}$ in Fig. 6. We determine the {\it effective}, size--dependent
critical
exponent $(\gamma/\nu)^z_{eff}(L)= d(\ln \chi^z)/d(\ln L)$
of the height of the peak in the staggered longitudinal
susceptibility $\chi^z_{st}$ for the simulated
system sizes at various fixed temperatures close to the
boundary of the AF phase. One observes a pronounced increase
of the effective exponent in a small interval of temperatures. This
behavior signals the change from a continuous
transition of Ising
type, with $(\gamma/\nu)^z \approx 1.9635$ \cite{PV}, to a transition
of first order, where the
asymptotic exponent is expected to be 3. A pronounced increase occurs
simultaneously, within the temperature resolution of the
simulations, for the effective exponent of the staggered transverse
susceptibility \cite{Ban}. Accordingly, it indicates a transition of first
order between the AF and SF phases. The ratio of the two
staggered susceptibilities exhibits an extremum which height is largely
independent of system size in the vicinity of the
multicritical point. Such a behavior seems to be consistent
with a bicritical Heisenberg point.

These observations and interpretations on the character of the
multicritical point have to
be viewed with care, having in mind controversial renormalization group
arguments. Early renormalization group calculations in
one loop order suggested a bicritical Heisenberg
point \cite{FN,KNF}. Later
work, in five loop order, found the biconical
point to be the stable
one \cite{CPV,PV2}, however, not ruling out a triple
point, at which at least one of the transition lines to the
paramagnetic phase is, eventually close to the multicritical
point, of first order. A most recent analysis indicates that
all three scenarios are possible, depending on the model
parameters \cite{Folk}. In our case, $\Delta= 0.8$, the
Monte Carlo simulations seem to favor a bicritical point, with
Ising and XY lines of the paramagnetic phase meeting the
line of first order between the AF and SF phases. Of course, a
crossover to a different multicritical behavior in
the immediate vicinity of the multicritical point may
happen. Simulational clarification of this aspect may need an enormous
amount of computer time.

\begin{figure}[b]
  \includegraphics[width=.7\textwidth]{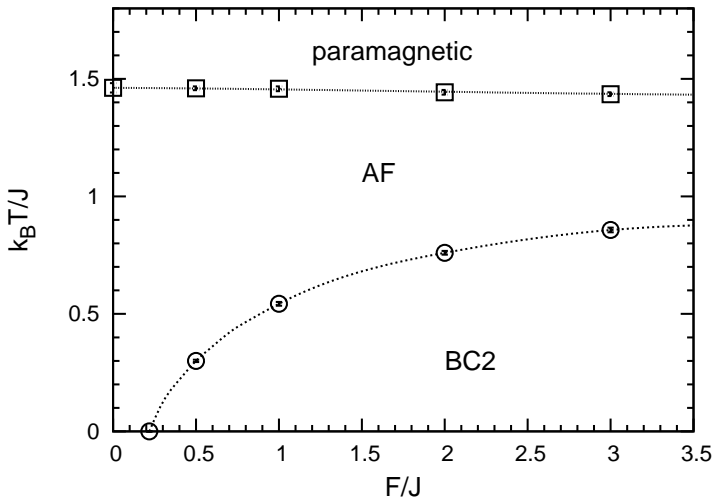}
  \caption{Phase diagram of the XXZ model, $\Delta= 0.8$, with a
    positive cubic anisotropy $F$ at fixed field, $H/J$ =1.8.
           }
  \label{eps7}
\end{figure}

Let us now turn to the Monte Carlo results for the full Hamiltonian,  
${\cal H}_{\mathrm{f}}$= ${\cal H}_{\mathrm{XXZ}}$ + ${\cal H}_{\mathrm{CA}}$, 
eqs. (1) and (2), including exchange and cubic anisotropies. We
simulated thermal properties for a few selected cases, where
discretized biconical structures play an important role. As
before, we set the exchange anisotropy $\Delta$ equal to 0.8.

For positive cubic anisotropy, $F >0$, favoring orientations of the
spins along the diagonals of the cubic lattice, a
discretized ordered BC2 phase may arise from the corresponding 
ground state, see Fig. 2. This behavior is illustrated in Fig. 7, when
fixing the field, $H/J$= 1.8, and varying $F$. At small values of 
$F$, there is an AF ordering at low temperatures. Above a critical 
value of $F$, $F > F_c(H=1.8J)= 0.218..J$, the low--temperature phase
is, indeed, of BC2 type. In that case, increasing the
temperature at fixed $F$, one
goes from the BC2 to the AF phase before then entering the disordered
paramagnetic phase. The transitions between the two ordered phases and
between the AF and disordered phases have been
located from monitoring the staggered magnetizations
and susceptibilities, the
specific heat, and Binder cumulants. An example is
shown in Fig.8, where the
staggered longitudinal and transverse magnetizations
are displayed, as a function of temperature, at $F/J$= 1.0. At the
transition between the AF and BC2 phases, the absolute staggered 
transverse magnetization, $|m^{xy}_{st}|$, drops rather
sharply, with an
accompanying anomaly in the absolute staggered
longitudinal magnetization, $|m^z_{st}|$.  The
staggered longitudinal magnetization is the order parameter
of the AF phase, and
it decreases rapidly on approach to the paramagnetic phase.  

\begin{figure}[b]
  \includegraphics[width=.7\textwidth]{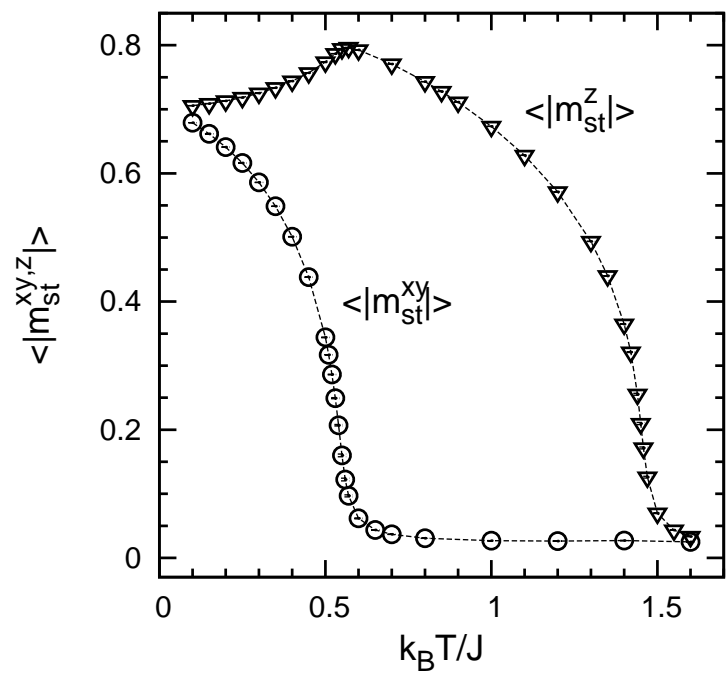}
  \caption{Absolute staggered longitudinal
    and transverse magnetizations as a function of temperature
    at fixed field, $H/J$= 1.8, and at fixed cubic
    anisotropy $F/J$= 1.0. Systems with $16^3$ sites have been
    simulated. Note that error bars are small compared to
    symbol sizes.
    }
  \label{eps8}
\end{figure}

As expected, the transition from the AF to the disordered phase
seems to be Ising like, as follows, e.g. from the critical
exponent describing the size--dependence of the height of the
peak in $\chi^z_{st}$. The 
transition from the BC2 to the AF phase is found to be consistent
with the XY universality class \cite{Ban}. Indeed, the transition
between the AF and
biconical phases has been predicted to belong to the
XY universality class in an early renormalization
group analysis \cite{BA}. Furthermore, the cubic anisotropy
is expected \cite{PV,CPV2} to be an irrelevant perturbation in the 
three--dimensional XY case. As depicted in Fig.7, there
is no multicritical point of the three phases
at the values of $F$ we studied.

A multicritical point may occur in
the $(H,T)$--plane, when fixing $F$, $F>0$. Indeed, our simulation 
results, especially at $F/J= 0.15$ and 1.5 \cite{Ban}, show
a phase diagram comprising the AF, BC2, SF, and paramagnetic
phases. However, much more extensive simulations, being
beyond the scope of the present study, would be needed to 
study the possible multicritical points in detail. Especially, the
order of the transition between the BC2 and SF phases deserves
a careful analysis. A transition of first order, as suggested
by the ground state analysis, would preclude a tetracritical
point \cite{BA}.
   
Finally, we briefly mention results for the case of negative
cubic anisotropy, $F<0$, favoring spin orientations along the 
cubic axes. As follows from the ground state
considerations, see Fig.2, biconical structures only show up at
relatively large negative values of $F$. Moreover, the tilt
angles usually seem to jump when going from the BC1 region to one of
its neighboring, i.e., ferromagnetic, AF, or SF, regions. 

Our Monte Carlo simulations, especially at $F/J$= -2 and varying the 
field at fixed temperatures \cite{Ban}, suggest that there
is an ordered BC1 phase which, however, gets destabilized at quite
low temperatures. Thorough analyses of full phase diagrams and the
types of the transition are well beyond the scope of the present work.  

\section{Summary}

Classical Heisenberg antiferromagnets on cubic lattices with uniaxial
exchange, $\Delta$, and cubic, $F$, anisotropies in a
magnetic field, $H$, have been 
studied, using ground state considerations and Monte Carlo techniques.

In addition to antiferromagnetic (AF) and
spin--flop (SF) structures non--collinear, biconical (BC) 
structures are observed, depending on
the strength of the anisotropies, the field,
and temperature.  They may occur as ground states, which, at
low temperatures, may contribute to thermal fluctuations of BC type
or may give rise to ordered BC phases.

Specifically, in the XXZ
antiferromagnet ($F$= 0) biconical structures are present as 
degenerate ground states at the critical field separating
the AF and SF states. At low temperatures near the transition
between the AF and SF phases, BC fluctuations show up, but they do
not lead to a distinct phase, in contrast, for
instance, to the situation for the XXZ magnet on a square
lattice, where the AF and SF phases are separated by a
narrow disordered phase. Moreover, in this study, the 
multicritical point of the three--dimensional XXZ
model, at which the AF, SF, and paramagnetic phases meet, has
been located
accurately for $\Delta$= 0.8. That point seems to have 
bicritical character, being the intersection
of two continuous transition lines between the AF and SF phases
to the paramagnetic phase with a transition line of first--order
between the AF and SF phases. Note that we
did not rule out a crossover to a different character in the
immediate vicinity of the multicritical point.

The cubic anisotropy $F$ may lead, depending on its sign, to
a tendency of the spin orientations along the diagonals of
the lattice, $F>0$, or along the cubic axes, $F<0$. In both
cases, BC as well as SF  structures have now discretized preferred
orientations of the $xy$--components of the spins, thereby 
breaking the rotational symmetry of the XXZ model.

For positive cubic anisotropy, $F >0$, biconical configurations
occur as ground states next to the SF and AF phases, giving rise to an
ordered BC2 phase. The transition
to the AF phase is found to be consistent with being in
the XY universality class, where the cubic anisotropy is
expected to be an irrelevant perturbation. Phase diagrams with
three ordered phases, AF, SF, and BC2, and a paramagnetic phase
have been determined.

At small negative cubic anisotropy, $F<0$, there are no 
biconical ground states. However, such structures, of
BC1 type, may be stabilized at larger negative values of $F$, between
the ferromagnetic and the AF or SF ground states. The biconical
ground states then give rise to an ordered BC1 phase at sufficiently
low temperatures.

{\bf Acknowledgements}

We should like to thank Martin Holtschneider, David Landau, and
David Peters for useful discussions. We also thank Reinhard Folk for
very helpful conversations and information on pertinent  
calculations prior to publication.


\begin{thebibliography}{00}

\bibitem{Neel} L.\ N\'eel, Ann. Phys.-Paris {\bf 5}, 232 (1936).
\bibitem{Gorter} C.\ J.\ Gorter and T.\ van Peski-Tinbergen,
  Physica (Utr.) \textbf{22}, 273 (1956).
\bibitem{LanBin} D.\ P.\ Landau and K.\ Binder, Phys.\ Rev.\ B\
  \textbf{17}, 2328 (1978).
\bibitem{Mourit} O.\ G.\ Mouritsen, E.\ K.\ Hansen, and S.\ J.\ K.\
  Jensen, Phys.\ Rev.\ B\
  \textbf{22}, 3256 (1980).
\bibitem{FN} M.\ E.\ Fisher and D.\ R.\ Nelson, Phys. Rev. Lett. {\bf
    32}, 1350 (1974).
\bibitem{KNF} J.\ M.\ Kosterlitz, D.\ R.\ Nelson, and  M.\ E.\ Fisher, Phys. Rev. B {\bf 13}, 412 (1976).
\bibitem{CPV} P.\ Calabrese, A.\ Pelissetto, and E.\ Vicari,
  Phys.\ Rev.\ B\ \textbf{67}, 054505 (2003).
\bibitem{Folk} R.\ Folk, Yu.\ Holovatch, and G.\ Moser,
  Preprint (2008).
\bibitem{HWS} M.\ Holtschneider, S.\ Wessel, and W.\ Selke,
  Phys.\ Rev.\ B\ \textbf{75}, 224417 (2007).
\bibitem{HS1} M.\ Holtschneider and W.\ Selke,
  Phys.\ Rev.\ B\ \textbf{76} (R), 220405 (2007).
\bibitem{HS2} M.\ Holtschneider and W.\ Selke,
  Eur.\ Phys. \ J.\ B\ \textbf{62}, 147 (2008).
\bibitem{Zhou1} C.\ Zhou, D.\ P.\ Landau, and T.\ C.\ Schulthess,
  Phys.\ Rev.\ B\ \textbf{74}, 064407 (2006).
\bibitem{HSL} M.\ Holtschneider, W.\ Selke, and R.\ Leidl,
  Phys.\ Rev.\ B\ \textbf{72}, 064443 (2005).
\bibitem{MT} H.\ Matsuda and T.\ Tsuneto, Prog.\ Theor.\ Phys.\ Supplement
  \textbf{46}, 411 (1970).
\bibitem{LF} K.-S.\ Liu and M.\ E.\ Fisher,
  J.\ Low.\ Temp.\ Phys.\ \textbf{10}, 655 (1973).
\bibitem{Kef} F.\ Keffer, in {\it Handbuch der Physik}, ed. by
  S. Fl\"ugge (Springer, Berlin, 1966), Vol. XVIII, Pt.2, p.1.
\bibitem{Aha} A.\ Aharony, Phys.\ Rev.\ B \textbf{8}, 4270 (1973).
\bibitem{Weg} F.\ Wegner, Solid State Commun.\ \textbf{12}, 785 (1973).
\bibitem{BA} A.\ D.\ Bruce and A.\ Aharony,
  Phys.\ Rev.\ B \textbf{11}, 478 (1975).
\bibitem{Aha2} A.\ Aharony,
  J.\ Stat.\ Phys.\ \textbf{110}, 659 (2003).
\bibitem{DJ} L.\ J.\ de Jongh (ed.), {\it Magnetic properties of
    layered transition metal compounds}, (Kl\"uwer, Dordrecht, 1990). 
\bibitem{T} T.\ Thio, C.\ Y.\ Chen, B.\ S.\ Freer, D.\ R.\ Gabbe, H.\
  P.\ Jenssen, M.\ A.\ Kastner, P.\ J.\ Picone, N.\ W.\ Preyer, and
  R.\ J.\ Birgeneau, Phys.\ Rev.\ B \textbf{41}, 231 (1990).
\bibitem{BS} T.\ Kroll, R.\ Klingeler, J.\ Geck, B.\ B{\"u}chner,
  W.\ Selke, M.\ H{\"u}cker, and A.\ Gukasov, J.\ Magn.\ Magn.\ Mat.\
 \textbf{290}, 306 (2005).
\bibitem{RG} H.\ Rohrer and C.\ Gerber, Phys. Rev. Lett. {\bf
    38}, 909 (1977).
\bibitem{BO} C.\ C.\ Becerra, N.\ F.\ Oliveira, A.\ Paduan-Filho, W.\
  Figueiredo, and M.\ V.\ Souza, Phys.\ Rev.\ B \textbf{38}, 6887 (1988).
\bibitem{FK} G.\ P.\ Felcher and R.\ Kleb, Europhys.\ Lett.\ \textbf {36}, 455
  (1996).
\bibitem{OU} K.\ Ohgushi and Y.\ Ueda, Phys. Rev. Lett. {\bf
    95}, 217202 (2005).
\bibitem{OP} Z.\ W.\ Ouyang, V.\ K.\ Pecharsky, K.\ A.\
  Gschneidner, D.\ L.\ Schlagel, and T.\ A.\ Lograsso, Phys.\ Rev.\ B
  \textbf {76}, 134415 (2007).
\bibitem{Ban} G.\ Bannasch, Diploma thesis, RWTH Aachen (2008).
\bibitem{Bin} K.\ Binder, Z. Physik B \textbf{43}, 119 (1981).
\bibitem{Bar} M.\ N.\ Barber, in {\it Phase Transitions and Critical
    Phenomena}, ed. by C.\ Domb and J.\ L.\ Lebowitz (Academic
  Press, New York, 1983), Vol. 8.
\bibitem{VR} K.\ Vollmayr, J.\ D.\ Reger, M.\ Scheucher, and K.\
  Binder, Z.\ Physik B\ \textbf{91}, 113 (1993).
\bibitem{PV} A.\ Pelissetto and E.\ Vicari, Phys.\ Rep.\ \textbf{368}, 549 (2002).
\bibitem{Ha1} M.\ Hasenbusch, K.\ Pinn, and S.\ Vinti, Phys. Rev. B \textbf{59}, 11471 (1999).
\bibitem{Ha2} M.\ Hasenbusch and T.\ T\"or\"ok, J. Phys. A \textbf{32}, 6361 (1999).
\bibitem{CD} X.\ S.\ Chen and V.\ Dohm, Phys.\ Rev.\ E \textbf {70}, 056136
  (2004); V.\ Dohm, Phys.\ Rev.\ E \textbf {77}, 061128 (2008).
\bibitem{SeSh} W.\ Selke and L.\ N.\ Shchur, J.\ Phys.\ A: Math. Gen.\textbf{38}, L739 (2005).
\bibitem{PV2} A.\ Pelissetto and E.\ Vicari, Phys.\ Rev.\ \textbf{76}, 024436 (2007).
\bibitem{CPV2} J.\ M.\ Carmona, A.\ Pelissetto, and E.\ Vicari,
  Phys.\ Rev.\ B\ \textbf{61}, 15136 (2000).




\end{thebibliography}
\end{document}